\documentclass[epj]{svjour}
\usepackage[switch,pagewise]{lineno}
\usepackage{graphicx}
\usepackage{color}
\usepackage[usenames,dvipsnames,svgnames,table]{xcolor}
\begin{document}
\title{Measurement of the cosmic ray energy spectrum using hybrid events of the Pierre Auger Observatory}
\author{Mariangela Settimo\inst{1,}\inst{2} for the Pierre Auger Collaboration\inst{3,*}}
%
%
\institute{Universit\"at Siegen, Siegen, Germany \and Dipartimento di Matematica e Fisica ``Ennio De Giorgi'', Universit\`a del Salento, Lecce, Italy \and Observatorio Pierre Auger, Malarg\"ue, Argentina}
\date{version 1.0 - 12.06.2012}
\small{  }
%

\abstract{
The energy spectrum of ultra-high energy cosmic rays above 10$^{18}$~eV is measured using the hybrid events collected by the Pierre Auger Observatory between November 2005 and September 2010. The large exposure of the Observatory allows the measurement of the main features of the energy spectrum with high statistics. Full Monte Carlo simulations of the extensive air showers (based on the CORSIKA code) and of the hybrid detector response are adopted here as an independent cross check of the standard analysis~\cite{spectrum2010}. The dependence on mass composition and other systematic uncertainties are discussed in detail and, in the full Monte Carlo approach, a region of confidence for flux measurements is defined when all the uncertainties are taken into account. 
An update is also reported of the energy spectrum obtained by combining the hybrid spectrum and that measured using the surface detector array. 
} 
\authorrunning{M. Settimo for the Pierre Auger Collaboration}
\titlerunning{Cosmic ray energy spectrum with the Pierre Auger Observatory}

\maketitle
\let\thefootnote\relax\footnotetext{*\emph{\footnotesize{Lists of authors and their affiliations appear at the end of the paper.}}}
\section{Introduction}
The features of the energy spectrum of ultra-high energy cosmic rays are intrinsically connected to the origin, nature and propagation of cosmic rays. 
At the highest energies, above 4$\times$10$^{19}$~eV, a suppression of the flux has been observed by the HiRes experiment~\cite{hires_spectrum}, the Pierre Auger Observatory~\cite{spectrum2010,SDspectrum,spectrum2011} and the Telescope Array~\cite{TA_spectrum}. This suppression is compatible with the predicted Greisen-Zatsepin-Kuz'min (GZK) effect~\cite{gzk1,gzk2}, even if other possibilities (\emph{e.g.} limits in the maximum energy at the source) cannot be excluded. A break in the power law spectrum, named  the ``ankle'', has also been reported around 10$^{18.6}$~eV~\cite{spectrum2010,hires_spectrum,spectrum2011,TA_spectrum,linsley,ankleexp,akeno,flyseye}. 
This feature is traditionally explained as the intersection of a steep Galactic component~\cite{linsley,hillas,ankle1,ankle2,ankle3} with a flatter extragalactic one though in this case the galactic component must extend up to energies above 10$^{18}$~eV, requiring a modification of the simple rigidity model of the cosmic ray confinement in the galaxy~\cite{knee}. 
 Other models explain the ankle structure as the distortion of a proton dominated spectrum through e$^{+}$/e$^{-}$ pair production of protons with the photons of the cosmic microwave background~\cite{hillas2,blumenthal,dip1,dip2}. A measurement of the cosmic ray flux in this energy range together with the knowledge of the mass composition over a wide energy range~\cite{massAuger1,massAuger2,massHiRes,massHiResMIA,massYakutsk1,massYakutsk2,massTA} may help to confirm these results and constrain different model scenarios.

The energy spectrum above 2.5~$\times$10$^{18}$~eV has been derived using data from the surface detector array of the Pierre Auger Observatory~\cite{SDspectrum}. This measurement has been extended to 10$^{18}$~eV~\cite{spectrum2010,spectrum2011} using the hybrid events, that are simultaneously observed by the fluorescence telescopes and by the surface detector. 

In this paper the measurement of the energy spectrum is updated to September 2010 and a method using detailed simulations of the extensive air showers (EAS) and of the hybrid detector response has been also developed. Hereafter we refer to this approach as ``full Monte Carlo''. It provides a complete treatment of the shower-to-shower fluctuations and an independent validation of the standard method (``fast simulations'') used in~\cite{spectrum2010} and described in detail in~\cite{exposure}. The standard method allows one to simulate a huge amount of events and to apply stricter analysis cuts which reduce the systematic uncertainties on the spectrum measurement. The two approaches adopted in this paper differ in the EAS and detector simulations and in the selection of events. Their advantages and drawbacks are discussed and the systematic uncertainties related to mass composition, hadronic interaction models and efficiency of the detector are studied.
 
The paper is organized as follows. In section~\ref{sect:auger} we introduce the Pierre Auger Observatory~\cite{augerNIM} and the hybrid detection mode, discussing its performance. The successive sections describe the steps to derive the energy spectrum, including the on-time and the hybrid exposure calculation using both the full Monte Carlo and the fast simulation approaches. The latter is introduced to provide larger simulation statistics in combination with a more strict analysis aiming to reduce the systematic uncertainties on the energy spectrum derived in section~\ref{sect:spectrum}.

\section{The Pierre Auger Observatory}\label{sect:auger}
 
The Pierre Auger Observatory~\cite{augerNIM}, located near Malarg\"ue, in the Province of Mendoza, Argentina, was designed to investigate the origin and the nature of ultra-high energy cosmic rays and has been taking data continuously since January 2004. 
It consists of an array of about 1600 water Cherenkov Surface Detectors (SD)~\cite{SDpaper} deployed over a triangular grid of 1.5~km spacing and covering an area of 3000~km$^{2}$. Each station is filled with 12~tons of purified water and three 9-inch photomultipliers (PMTs) detect the Cherenkov light produced in water by the secondary particles of the air shower.  An SD event is formed when at least 3 non-aligned stations selected by the local station trigger are in spatial and temporal coincidence. 
The ground array is overlooked by 24 fluorescence telescopes, grouped in four sites, making up the fluorescence detector (FD)~\cite{FDpaper}. The FD observes the longitudinal development of the shower in the atmosphere (\emph{i.e.} the energy deposit as a function of the atmospheric depth) by detecting the fluorescence light emitted by excited nitrogen molecules and Cherenkov light~\cite{giller_cherenkov,nerling_cherenkov2} induced by shower particles in air. FD provides a calorimetric measurement of the primary particle energy, only weakly dependent on theoretical models. Each telescope comprises a wide-angle segmented mirror, which reflects the UV light onto a camera of 440 PMTs (called hereafter ``pixels''). An FD event is identified by a sequence of triggered pixels. An event detected by at least one FD telescope and one SD station is called ``hybrid''. The combination of the timing information from the FD and the SD provides an accurate determination of the geometry of the air showers. Moreover the use of the hybrid detector allows one to cover the energy range below the threshold ($\sim$~10$^{18.5}$~eV) for the independent SD event trigger~\cite{SDpaper}.

A detailed description of the hybrid reconstruction is given in~\cite{FDpaper} and it can be summarized in 2 steps: (i) the determination of the shower detector plane (SDP) which is the plane containing the FD telescope and the shower axis; (ii) the determination of the shower axis within the SDP. 
The pointing directions of the pixels allow the reconstruction of the SDP with a typical uncertainty of about a few tenths of a degree. Then the timing information of the pixels is used to derive the arrival direction and the distance of the shower axis from the FD. In the hybrid reconstruction, this procedure is supplemented with the shower front arrival time at the SD station with the largest signal. The arrival direction of the primary particle and the impact point of the shower at the ground are thus determined with a resolution of about 0.6$^{\circ}$ and 50~m, respectively, above 10$^{18}$~eV. 

Once the geometry is known, the longitudinal profile of the shower is reconstructed taking into account the light hitting the camera and the scattering and absorption of fluorescence and Cherenkov light during its propagation to the FD as discussed in~\cite{unger_energydep}. The electromagnetic energy released by the shower in the atmosphere is obtained by fitting the longitudinal profile to a Gaisser-Hillas function~\cite{GaisserHillas} and then integrating over the range of atmospheric depth. The total energy of the primary particle is derived by correcting for the invisible energy~\cite{barbosa} carried by penetrating particles, which weakly depends on the primary type and on the hadronic interaction models.

The interest in the energy range around and below 10$^{18}$~eV has motivated several upgrades of the detectors: (i) the Auger Muon and Infill for the Ground Array (AMIGA)~\cite{amiga,amiga2} which consists of buried muon counters and of a dense array (``infill'') of water Cherenkov stations deployed on a grid of 750~m; (ii) three fluorescence telescopes (High Elevation Auger Telescopes, HEAT)~\cite{heat} installed with elevation angles between 30$^{\circ}$ and 60$^{\circ}$. Both the infill array and the HEAT telescopes are complete and are now taking data. In the near future they will allow the extension of the energy spectrum below the current energy threshold (10$^{18}$~eV) with high statistics. 

\section{ Exposure of the hybrid detector}\label{sect:exposure}
\begin{figure}[!t]
\hspace{-0.4cm}
 \resizebox{0.55\textwidth}{!}{\includegraphics{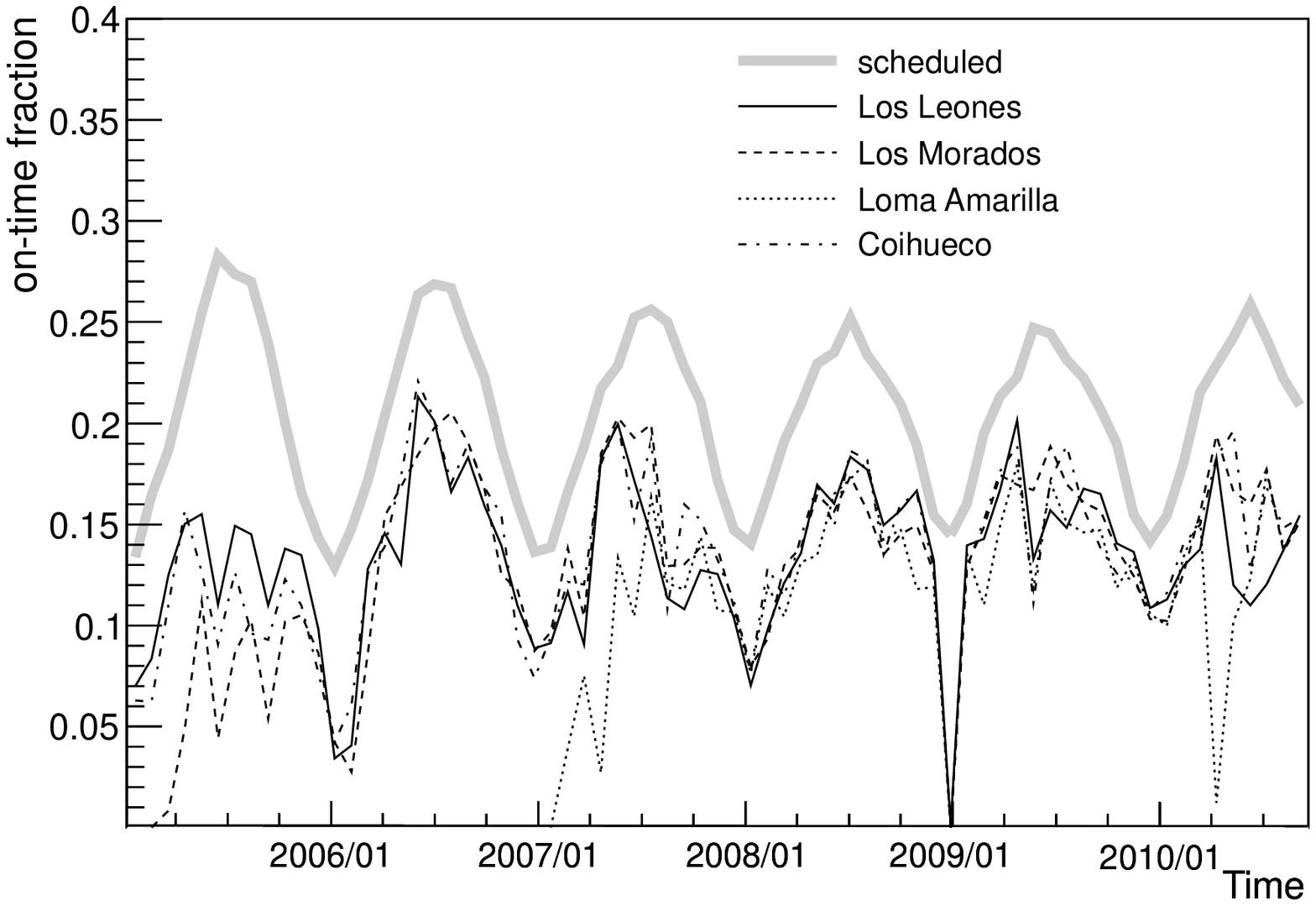}}
\hspace{-0.8cm}
 \resizebox{0.55\textwidth}{!}{\includegraphics{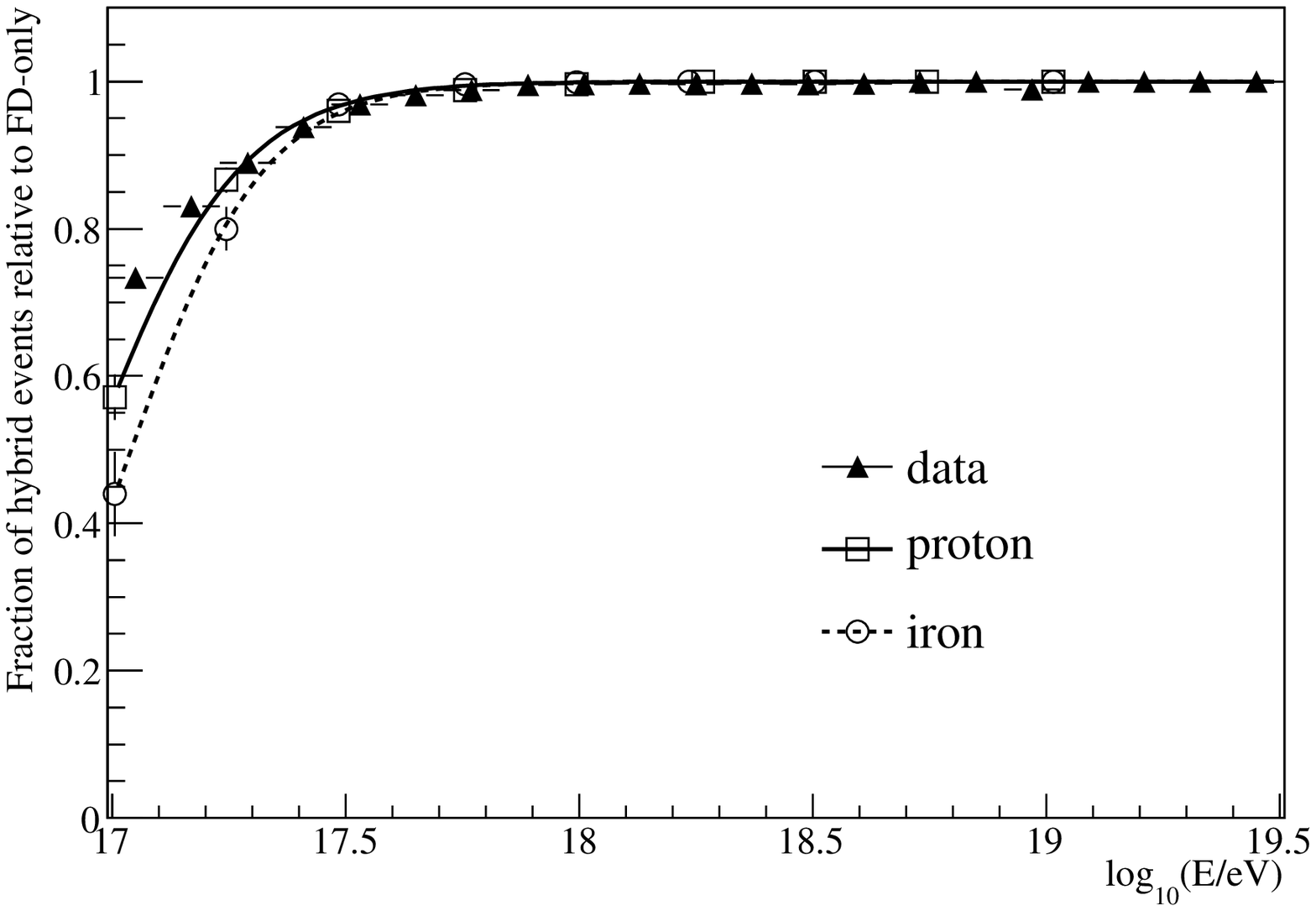}}
\caption{Left: time evolution of the average hybrid on-time fraction for the four FD sites (thin lines). The seasonal modulation, the starting of commissioning phases of the FD and temporary failures are visible. The gray line represents the scheduled data-taking time fraction. It is currently limited to the nights with a moon-fraction lower than 60\%. Right: relative hybrid trigger efficiency from full Monte Carlo simulation for proton and iron primaries. An estimate of the hybrid trigger efficiency calculated using data is also shown for comparison.}\label{fig:Ontime}
\end{figure}
The differential flux $J(E)$ of cosmic rays is defined as the number (d$N_{\rm{inc}}$) of events incident on the surface element d$S$ and solid angle d$\Omega$, in the time interval d$T$ and energy bin d$E$: 
\begin{equation}
  J(E)  =  \frac{\textrm{d}^4N_\mathrm{inc}}{\textrm{d}E \textrm{d}S \textrm{d}\Omega \textrm{d}t} \simeq \frac{\Delta N_\mathrm{sel}(E)}{\Delta E}\frac{ 1}{\mathcal{E}(E)};
  \label{eq.UHECRflux}
\end{equation}
where $\Delta N_\mathrm{sel}(E)$ is the number of selected events in the energy bin centered in $E$ and having a width $\Delta E$, and $\mathcal{E}(E)$ is the exposure of the detector, defined as: 
\begin{equation}
\mathcal{E}(E)  = \int_{T}\int_{\Omega}\int_{S} \varepsilon(E,t,\theta,\phi,x,y) ~ \cos \theta ~ \textrm{d}S ~ \textrm{d}\Omega ~ \textrm{d}t
\label{eq:exposure}
\end{equation}
$\varepsilon$ is the overall efficiency, including detection, reconstruction and selection of the events and the evolution of the detector in the time period $T$, $\theta$ and $\phi$ are the zenith and azimuth angles respectively, with $0^\circ<\theta<60^\circ$ and $-180^\circ<\phi<180^\circ$, $\Omega$ and $\textrm{d}\Omega=\textrm{d}\cos\theta\,\textrm{d}\phi$ are the total and  differential solid angles, $\cos\theta \textrm{d}S$ is the differential projected surface element. The area $S$ encloses the full detector array and is chosen sufficiently large to ensure a negligible trigger efficiency outside of it.
During part of the time period analysed here, the Observatory was still under construction. The last FD site was completed at the beginning of 2007 while the last SD station was deployed in June 2008. Also in a steady configuration, the status of each detector may change due to temporary hardware failures, maintenance, connections problems, etc. Moreover, the data taking and the trigger efficiency for the fluorescence detection depends on the sky and weather conditions (lunar cycle, brightness of the sky, lightning, wind, cloud coverage and aerosol content). These varying configurations have to be reproduced in simulations for a correct determination of the exposure.

\subsection{On-time calculation}\label{sect:ontime}

The calculation of the on-time for each FD telescope is derived by taking into account the status of the data acquisition, of the telescopes, pixels, communication system, etc. Details of the on-time and exposure calculations are given in~\cite{exposure}. Since July 2007 a new tool based on the monitoring system~\cite{monitoring} has been developed for the on-time calculation, accounting for several terms as discussed below. Before this date, the information on the status of the detector was extracted from a minimum bias datastream which includes sub-threshold FD events, recorded at a rate about 8 times higher than the standard one. The on-time fractions derived using these two tools have been compared in a common time window and they agree to within 3-4\%.

As a compromise of accuracy and stability, the on-time of the hybrid detector is calculated in temporal bins, $\Delta t$, of 10 minutes. In each time bin $t$, the detector on-time $f(i,t)$ for the telescope $i$ ($1\leq i \leq 24$) and FD site $s$ is given by:
\begin{equation}
  f(i,t) = \varepsilon_{\mathrm{shutter}}(i,t) \cdot (1-T^{dead}(i,t)) \cdot \epsilon_{\mathrm{CDAS}}(s,t)
\label{eq:ontime}
\end{equation}
where $\epsilon_{\mathrm{CDAS}}$ refers to the status of the Central Data Acquisition System (CDAS), including connection failures between the SD, the FD and the radio communication towers, $\varepsilon_{\mathrm{shutter}}(i,t)$ gives the fraction of time in which the shutters of each telescope are opened and $T^{dead}(i,t)$ is the cumulative dead time for each telescope divided by $\Delta t$. The latter is mostly related to the finite readout speed of the DAQ system, to buffer overflows, vetoed time intervals induced by the operation of the LIDAR system~\cite{lidar} and vetoes from the CDAS in the case of an excessive rate of FD triggers (\emph{e.g.}, because of lightning).

In Fig.~\ref{fig:Ontime} (left) the monthly averaged on-time fraction is shown for each FD site (thin lines) as a function of time. The duty-cycle for the FD mainly depends on moon-cycle (the expected mean value is plotted as a gray bold line) and seasonal changes in the daylight and darkness durations. Data taking is currently limited to nights with a moon-fraction smaller than 60\%. Compared to the nominal value of darkness, the hybrid detector is in acquisition for about 80-85\% of time, which includes good weather conditions (reasonable brightness of the sky, cloud coverage, wind and no rain) and detector efficiency. In the time interval considered for this analysis, the average overall duty-cycle for the FD is about 14\%, over a typical year with stable DAQ conditions.

The status of each SD station is monitored by the CDAS every second. This information is used in simulations to reproduce the actual status of the array. Moreover, time periods with trigger related problems are excluded from the analysis~\cite{SDpaper}. 
 Systematic uncertainties in the hybrid on-time are estimated to be about 4\% based on a cross-check performed using laser shots from the Central Laser Facility (CLF)~\cite{aerosol}.

\subsection{Time dependent simulations with the full Monte Carlo approach}\label{sect:corsika}
\begin{figure}[tb]
\hspace{-0.4cm}
\resizebox{0.55\textwidth}{!}{\includegraphics{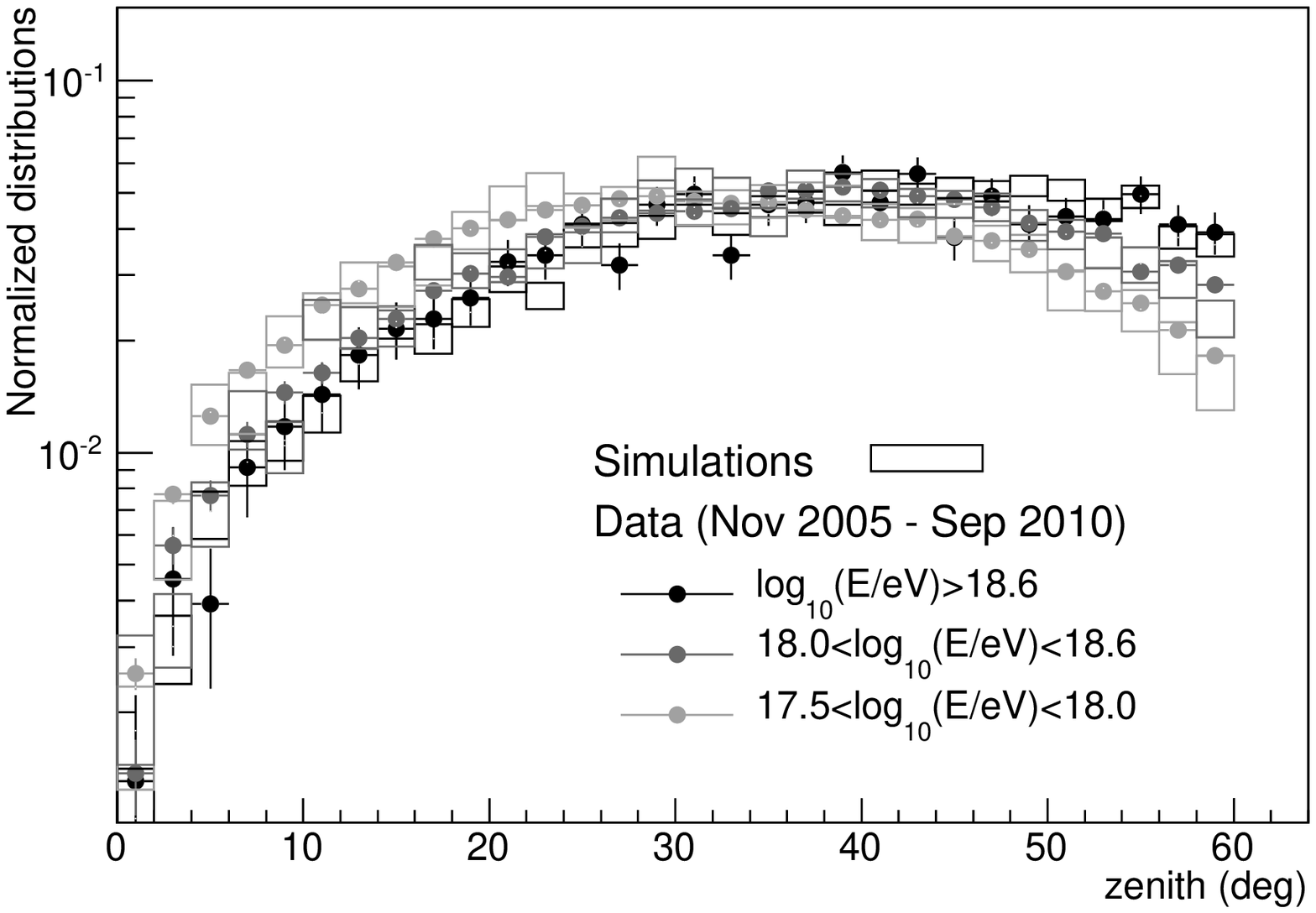}}
\hspace{-.8cm}
\resizebox{0.55\textwidth}{!}{\includegraphics{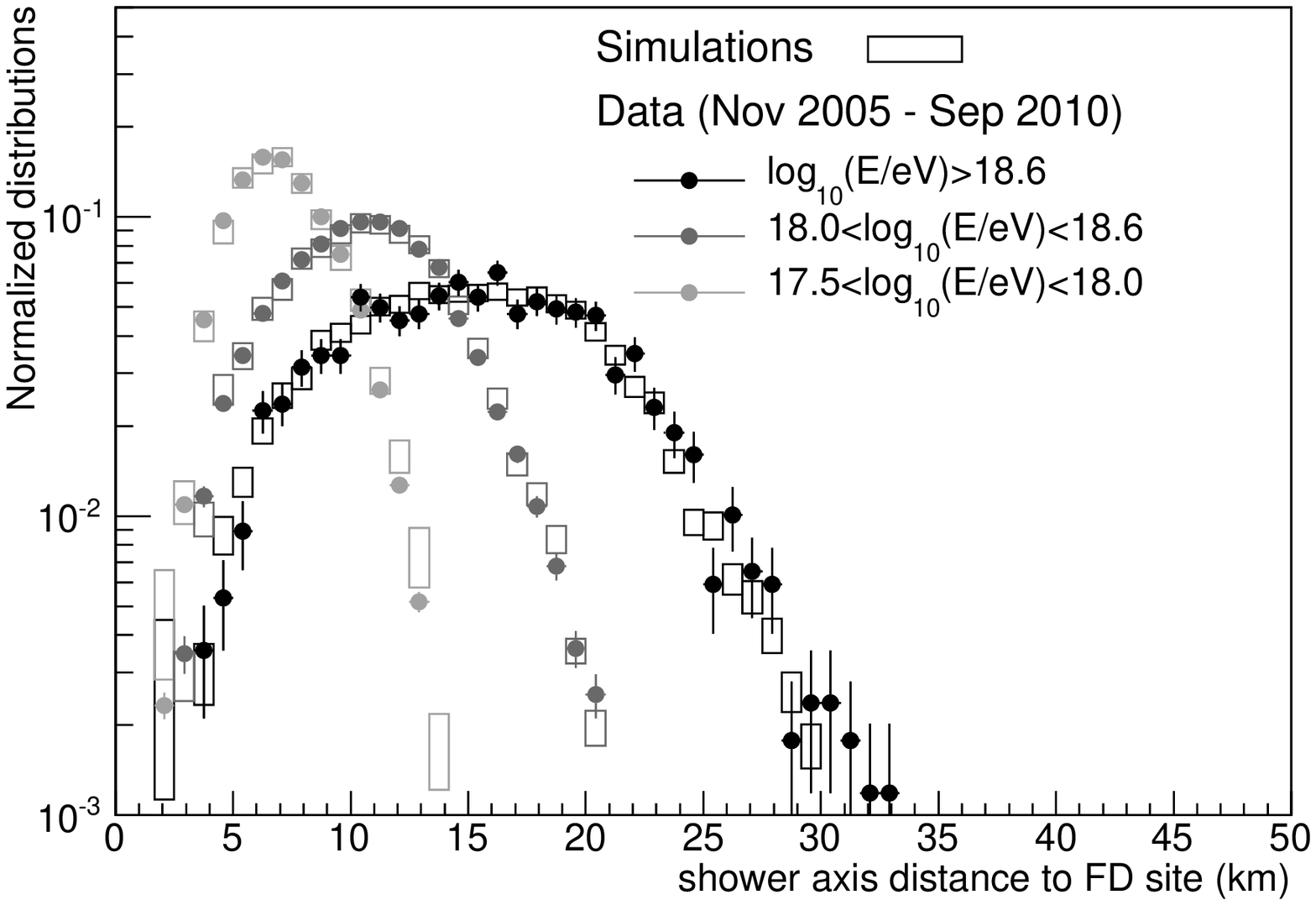}}
\caption{Examples demonstrating the agreement between simulation (boxes) and data (markers) in separate energy bins. For this purpose, the simulations are reweighted according to the spectral index given in~\cite{spectrum2010} and a mixed (50\% proton - 50\% iron) composition is assumed. The zenith angle and the distance of the shower axis to the FD are shown on the left and right, respectively. \label{fig:dataMC}}

\end{figure}
For an accurate determination of the exposure, all the detector configurations are taken into account by performing detailed time-dependent simulations of the air shower development and of the detector response. Air showers are simulated using the CORSIKA~6.960~\cite{corsika} Monte Carlo code which provides the longitudinal profile of the showers as well as the secondary particles at the ground (\emph{i.e.} at the Pierre Auger Observatory altitude). A sample of 70000 CORSIKA showers is used in the present analysis. The air showers have been generated with zenith angle $\theta$ distributed as $\sin\theta\cos\theta$ (with $\theta~<$~65$^{\circ}$), according to the projection on a surface detector of a isotropic flux of cosmic rays, and with energy ranging  between 10$^{17}$~eV and 10$^{19.5}$~eV, according to a power law spectrum (spectral index $\gamma = -1$) and in intervals of 0.5 in the logarithm of energy.  Simulations are performed using QGSJET-II.03~\cite{qgs2}, QGSJET-I~\cite{qgs1} and Epos-1.99~\cite{epos} as hadronic interaction models at high energy and FLUKA~\cite{fluka} at low energy. Moreover, seasonal models of the atmospheric conditions (pressure, temperature and air density) as measured in Malarg\"ue~\cite{MalargueAtmoProfile} are used in addition to the US standard model~\cite{USStandard}.

The hybrid detector response is simulated using the Auger Offline software~\cite{offline}. The FD simulation chain~\cite{FDsim} covers the physical processes involved in the fluorescence technique, such as the production of fluorescence and Cherenkov photons in the atmosphere, their propagation to the telescope, the ray-tracing of photons in the Schmidt optics, and the response of the electronics and multi-level trigger.
The secondary particles of the shower reaching the ground are injected and traced inside the SD stations and the detector response (including PMTs and electronics) is simulated with Geant4~\cite{geant4}.  
The positions of the impact point of the shower at the ground (hereafter briefly named the ``core'') are generated uniformly on the surface $S$ including the SD array plus an additional area surrounding the boundaries of the array to take into account events landing outside the array that may still be detected and successfully reconstructed. Its width is energy dependent  and its boundaries were defined according to the Lateral Trigger Probability (LTP) functions~\cite{LTP} as boundaries outside of which the trigger probability of a station is negligible (less than 0.5\% for events at 60$^{\circ}$). 
Fig.~\ref{fig:Ontime} (right) shows that every FD event above 10$^{18}$~eV is accompanied by at least one SD station, independently of the mass and direction of the incoming primary particle. For an accurate and unbiased measurement of the energy spectrum, the analysis discussed in this paper is limited to the energy range above 10$^{18}$~eV.
A dynamical resampling method has been implemented to optimize the usage of the CORSIKA showers: if an event has no chance of triggering the FD because it is too far from all FD sites, a new core position is generated until the event lands within a (energy dependent) maximum distance from FD, for which the trigger probability is not null~\footnote{We found that less than 1 out of 10$^{5}$ events trigger outside the maximum distance used for the dynamical resampling, even considering realistic atmospheric conditions and different fluorescence yield models.}. This shower resampling is quite useful at low energy because only events landing within a few kilometers from the FD telescope may have a chance to trigger FD. Within this triggerable region each CORSIKA shower is re-used 7 times, each time with a different core location and a different GPS time (\emph{i.e.} different detector condition and status), which ensure a negligible degree of correlation.
In the time-dependent simulations the actual status of each telescope and SD station, as well as realistic atmospheric conditions (transparency of air, aerosol content, etc.) are taken into account. 
The detailed simulation of the surface array enables the SD-event trigger to be independently formed, realistically reproducing the full acquisition system.

\subsection{Event selection and data/Monte Carlo comparison}\label{sect:eventselection}

A crucial aspect for the measurement of the energy spectrum is the accurate determination of the shower energy. High quality events with $\theta\,<$~60$^{\circ}$ and with a successful and good reconstruction of the arrival direction and of the longitudinal profile (Gaisser-Hillas fit with a $\chi^{2}/ndof\,<$~2.5) are selected. Moreover, we require that the depth, $X_{\rm{max}}$, corresponding to the maximum development of the shower, is observed, the fraction of Cherenkov light with respect to the overall signal detected by FD is smaller than 50\% and the uncertainty on the reconstructed energy is less than 20\%. 
These selection criteria ensure an average energy resolution of about 10\%, almost independent of energy (above $\sim$~10$^{17.5}$~eV).
For a precise energy estimation, the analysis uses only events with available information on the aerosol content~\cite{aerosol}. Since clouds may obscure or distort part of the longitudinal profile, the coverage measured by the Lidar system~\cite{lidar} is required to be lower than 25\%. A further cut is applied to reduce possible FD trigger effects induced by the systematic uncertainty on the energy scale, estimated to be at the level of 22\%~\cite{digiulio}. 
Events are selected if these are landing within a fiducial distance for which the FD trigger efficiency is flat within 5\% when shifting the energy scale by its systematic uncertainty.
The reliability of the quality criteria are checked by comparing the distributions of several observables taken from both data and Monte Carlo. Two examples are given in Fig.~\ref{fig:dataMC} for the zenith angle (left) and the shower axis distance to the FD (right). In both plots, the comparison is performed in three separate energy intervals and simulations are reweighted according to the spectral indices obtained in~\cite{spectrum2010}. The agreement between data (markers) and simulations (lines) is fairly good for both observables in the three energy ranges. 

\begin{figure}[!t]
\centering
\hspace{-0.4cm}
\resizebox{0.52\textwidth}{!}{\includegraphics{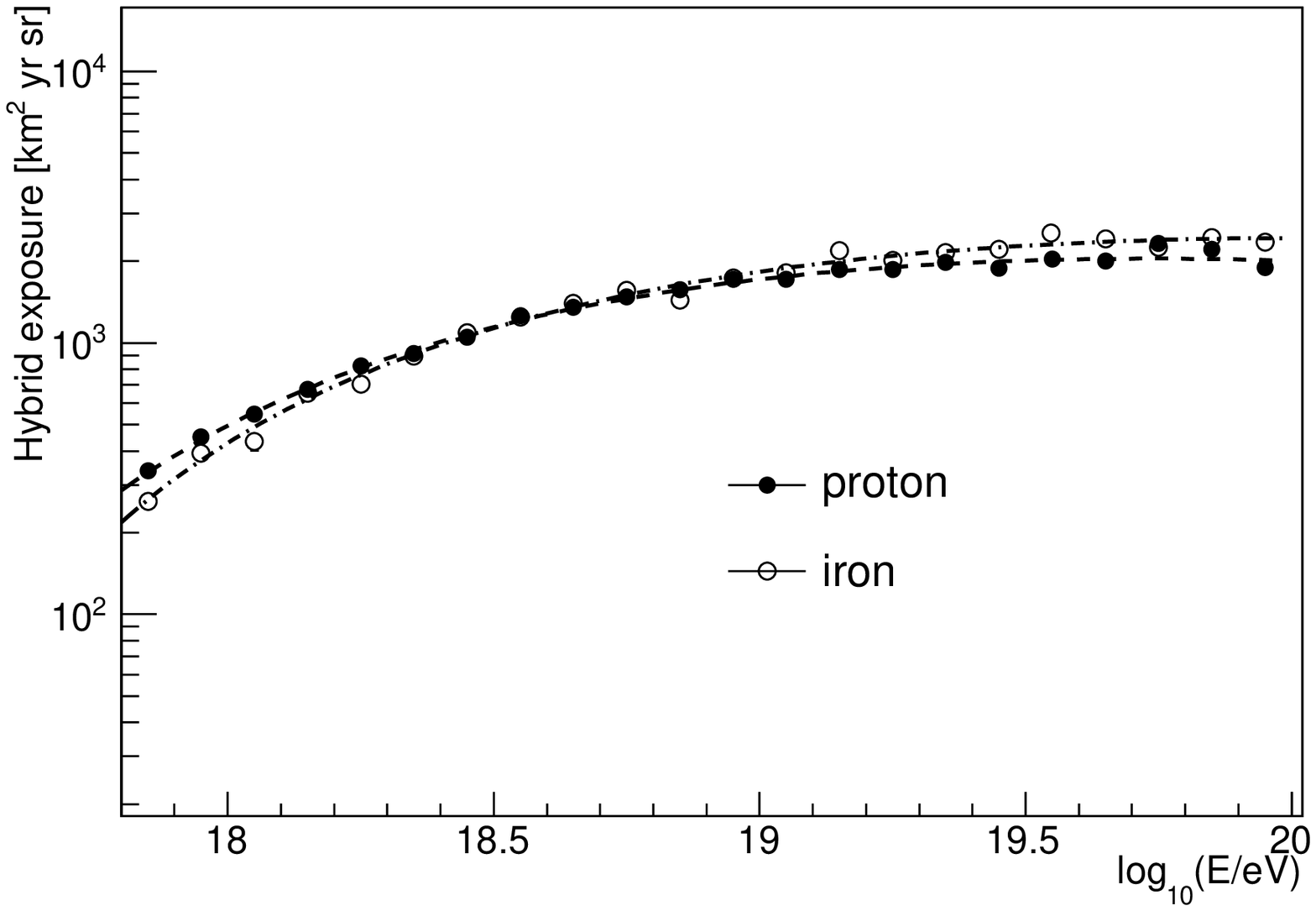}}
\hspace{-0.8cm}
\resizebox{0.52\textwidth}{!}{\includegraphics{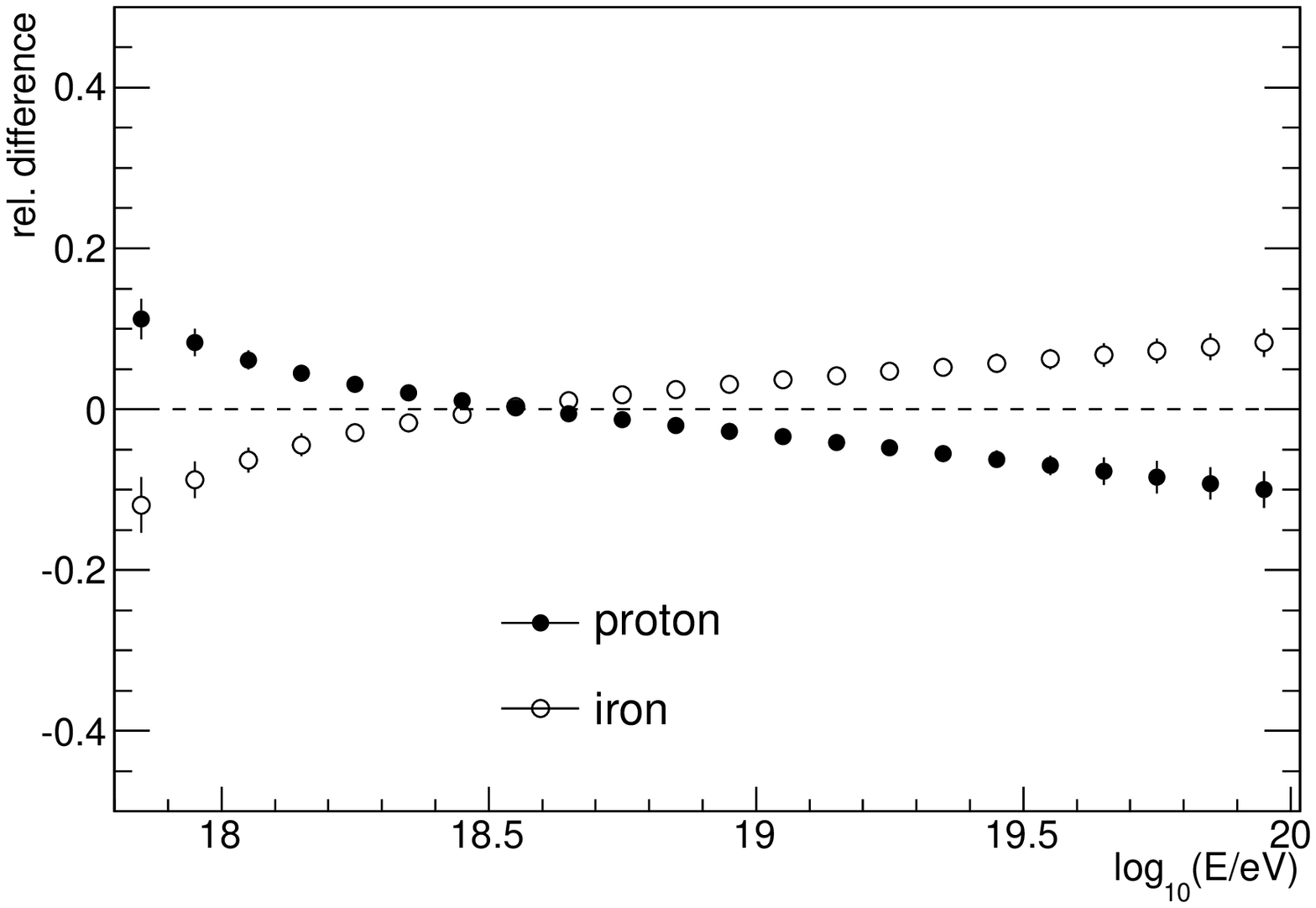}}
\centering
\caption{Left: hybrid exposure for proton (filled markers) and iron (empty markers) as a function of energy. Right: difference in the exposure for proton and iron, relative to that for a mixed (50\% proton - 50\% iron) composition.}\label{fig:exposureCORSIKA}
\end{figure}

The hybrid exposure, given by equation~\ref{eq:exposure}, is shown in Fig.~\ref{fig:exposureCORSIKA} for proton and iron primaries. A mass composition dependence is visible, particularly at low energies. Indeed, at these energies iron primaries, developing higher in the atmosphere, have a smaller probability of being detected and being well observed in the FD field of view (FOV) than protons. At higher energies, events far away from an FD are mainly selected. For these events the lower bound of the FD field of view disfavoures deep (\emph{i.e.} proton induced) showers. In Fig.~\ref{fig:exposureCORSIKA} (right), the ratio of the exposure of each pure composition relative to a mixed one (50\% proton - 50\% iron) is given as a function of energy. At energies above 10$^{18}$~eV, the difference is less than $\sim$~10\% depending on the energy and it rapidly increases at lower energy.
 As a consequence of this dependence, and given the lack of accurate knowledge of the nature of primary cosmic rays in this energy range, a mass-independent measurement of the energy spectrum cannot be performed. Estimates of the energy spectrum can be derived assuming a pure proton and a pure iron composition, which provides a confidence region (see section~\ref{sect:spectrum}) in which we expect the spectrum to be confined if the cosmic ray flux is dominated by nuclear primaries. Photons as primary particles at these energies are strongly constrained~\cite{photonlimits,SDphotons}. 

\subsection{Exposure calculation with fast simulations } \label{sect:conex}

\begin{figure}[!t]
\hspace{-0.5cm}
  \includegraphics[width=0.55\textwidth]{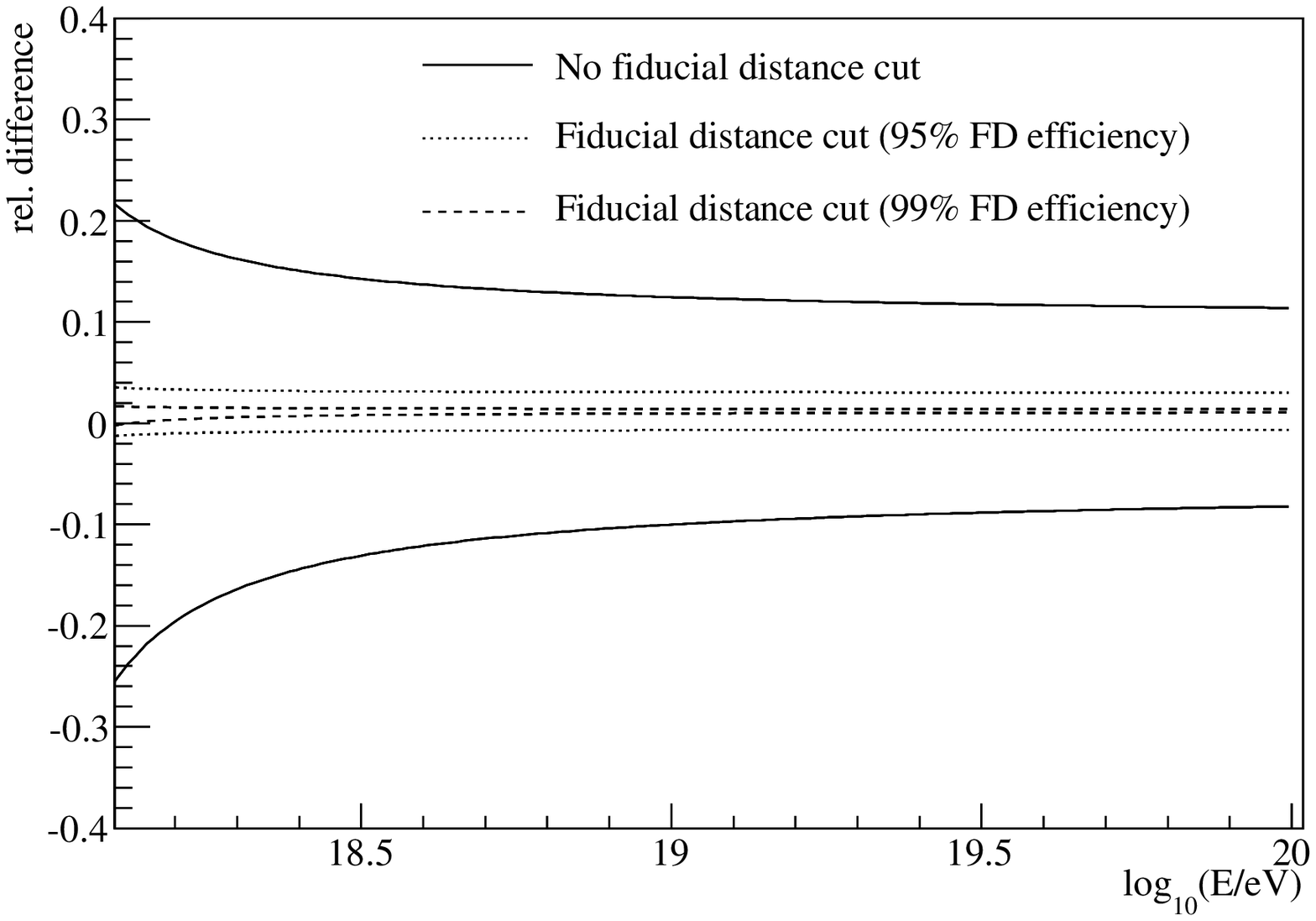}
\hspace{-0.9cm}
   \includegraphics[width=0.55\textwidth]{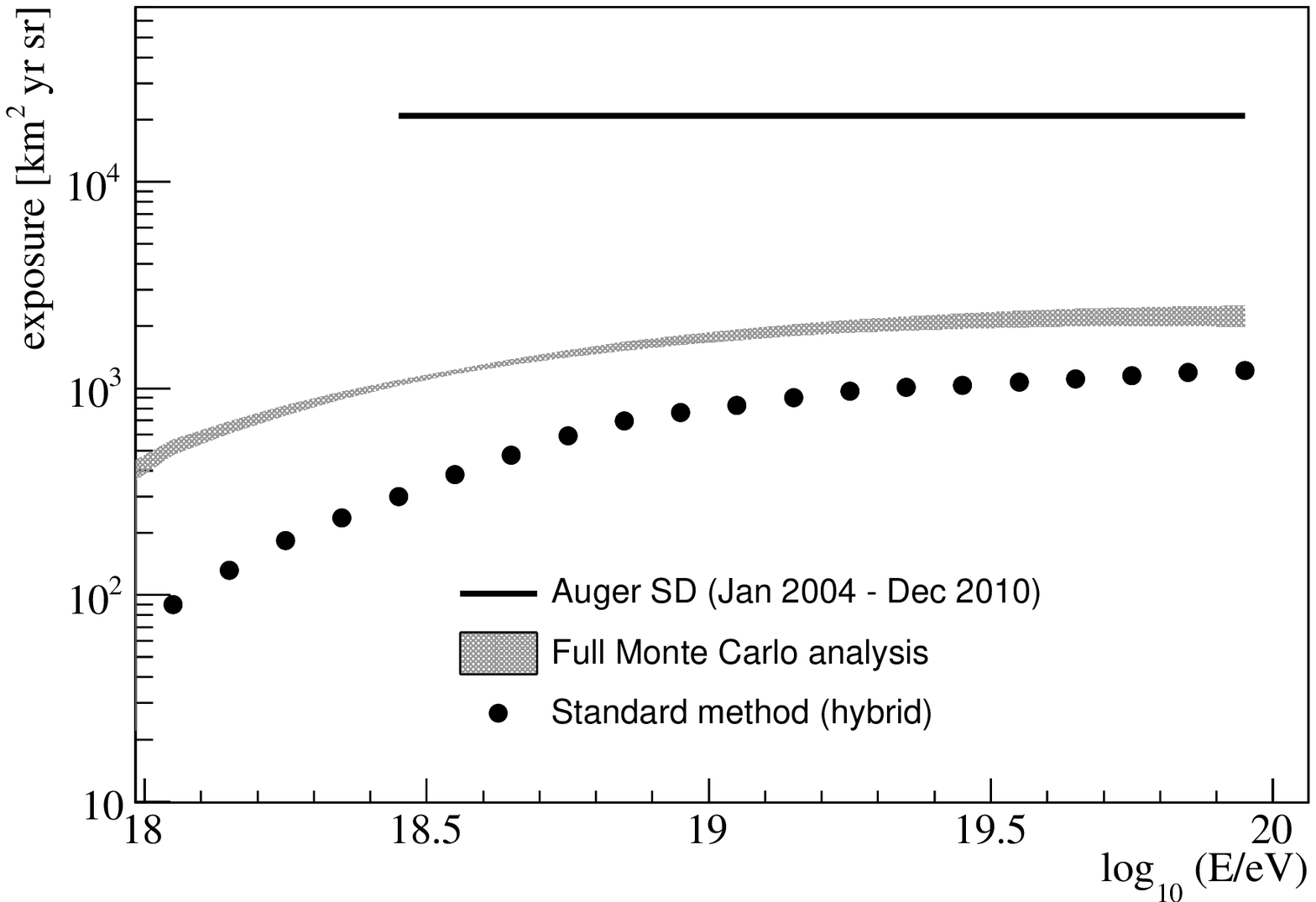}
\caption{Left: relative difference between the hybrid exposure with a $\pm$~22\% shift and the nominal value (solid lines). Fiducial distance cuts are designed to reduce this effect requiring an efficiency higher than 95\% (dotted line) and higher than 99\% (dashed line). 
Right: exposure calculated between November 2005 and September 2010 for the hybrid detector. The mixed composition assumption is plotted for the standard method, based on fast simulations and on an analysis designed to reduce the systematic uncertainties due to mass composition. For the full Monte Carlo approach, a shaded region delimited by the pure proton and pure iron cases is shown. Its higher value results from the different selection criteria (see text). The exposure for the SD array (line), between January 2004 and December 2010, is also given.}\label{fig:systematics}
\end{figure}

As discussed in the previous section, the limited field of view of the fluorescence detector and the requirement of observing the shower maximum may introduce a different selection efficiency for different primary masses. To reduce the impact of mass composition on the hybrid exposure, a dedicated analysis has been performed by defining a geometrical volume which guarantees comparable selection efficiency to all nuclear primaries. For a given energy and event geometry, this volume is defined by setting the lower and upper boundaries (expressed in atmospheric depth) of the FD field of view. This ``fiducial FOV cut'' is applied in addition to the quality selection criteria described in the previous section and it reduces the primary mass dependence to 8\% (1\%) at 10$^{18}$~eV (above 10$^{19}$~eV)~\cite{exposure}. Moreover the cut on the FD fiducial distance, introduced in section~\ref{sect:eventselection}, is applied here more strictly, requiring an FD efficiency larger than 99\% independently of a shift of $\pm$~22\% on the energy scale. The benefit of this cut is demonstrated in Fig.~\ref{fig:systematics}, left. The lines show the relative difference between the exposure with $\pm$22\% shifted energy and the nominal value, for two definitions of the fiducial distance cut (dashed and dotted) and if the cut is not applied (solid).  

Because of the strict selection and the demanding resources for a high statistics sample of full Monte Carlo simulations, a fast and simplified approach has been adopted to produce a large sample of simulations in a reasonable computational time. This method uses the CONEX~\cite{conex} code to simulate the air shower profile by a Monte Carlo generation of the first interactions and then a numerical solution of the cascade equations. This EAS generator is extremely fast and reproduces accurately the longitudinal profile including its shower-to-shower fluctuations~\cite{conex,fluctuations}. However it has the drawback of not providing the distribution of secondary particles at the ground and, consequently, the response of the detector cannot be directly simulated. The SD trigger is thus extracted using the Lateral Trigger Probability (LTP) functions, which parametrize the trigger probability of each SD station as a function of its distance to the shower axis, and of the energy and arrival direction of the primary cosmic ray. Even though the probability of having at least one station for each FD event is unity above 10$^{18}$~eV, this procedure may be relevant for low energy and inclined events.
The SD timing information needed in the hybrid reconstruction mode is modelled with an NKG-like~\cite{nkg1,nkg2} function for the lateral distribution of the air showers. The validity of this assumption has been verified in~\cite{exposure}.

The FD response is fully simulated with the approach and on-time tools described above. The hybrid exposure calculation based on the ``fast simulation'' and on the corresponding selection criteria, is the same used in~\cite{spectrum2010,spectrum2011} for previous spectrum measurements. For this reason, the approach discussed in this section is also referred to as the ``standard method''. The exposure is shown in Fig.~\ref{fig:systematics} (right) for a mixed composition (filled dots) of 50\% proton and 50\% iron primaries. This assumption, especially for energies above 10$^{19}$~eV, is well justified because of the reduced mass composition dependence of the exposure. The residual difference between proton and iron, for the standard method, is accounted as systematic uncertainty (see section~\ref{sect:systematics}). For comparison, the exposure derived in section~\ref{sect:corsika} (full Monte Carlo approach) is here plotted as a band delimited by the pure proton and pure iron assumptions. As a consequence of the less strict cuts, this exposure is significantly higher, especially at low energies. However the systematic uncertainty related to mass dependence is higher. The tighter analysis cuts introduced in this section have also been applied to full Monte Carlo approach and the derived exposure is in agreement with the one from the fast simulation. This check further validates the reliability of the standard method. As reference, the exposure derived using the SD-only array, valid for energies above $\sim$~10$^{18}$~eV, is shown in Fig.~\ref{fig:systematics} (right) until December 2010. Details on the SD exposure are given in~\cite{SDpaper,spectrum2011}.

\subsection{Systematic uncertainties}\label{sect:systematics}

For the standard method, the overall systematic uncertainty in the exposure calculations has been estimated as 10\% (6\%) at 10$^{18}$~eV ($>$ 10$^{19}$~eV). It includes the contributions listed below and discussed in detail in~\cite{exposure}. The uncertainties in mass composition (8\% at 10$^{18}$~eV and 1\% above 10$^{19}$~eV) and in the on-time calculation ($\sim$4\%) have been discussed in the previous sections. As a result of the checks with CLF laser shots and between SD data and the Monte Carlo simulations, the exposure has been reduced by 8\% to account for lost events and an upper limit to the remaining systematic uncertainty of 5\% has been derived. Different hadronic interaction models used for simulations may produce different predicted properties of the showers and consequently different trigger and selection efficiency. The impact on the exposure has been studied in~\cite{exposure} using QGSJETII-03 and Sibyll 2.1~\cite{sibyll} as hadronic interaction models and the average effect is lower than 2\% over the full energy range. Furthermore, an additional uncertainty of about 2\% is quoted due to the choice of the index of the input spectra used in simulations. 

\begin{figure}[!tb]
\hspace{-0.4cm}
 \includegraphics[width=0.51\textwidth]{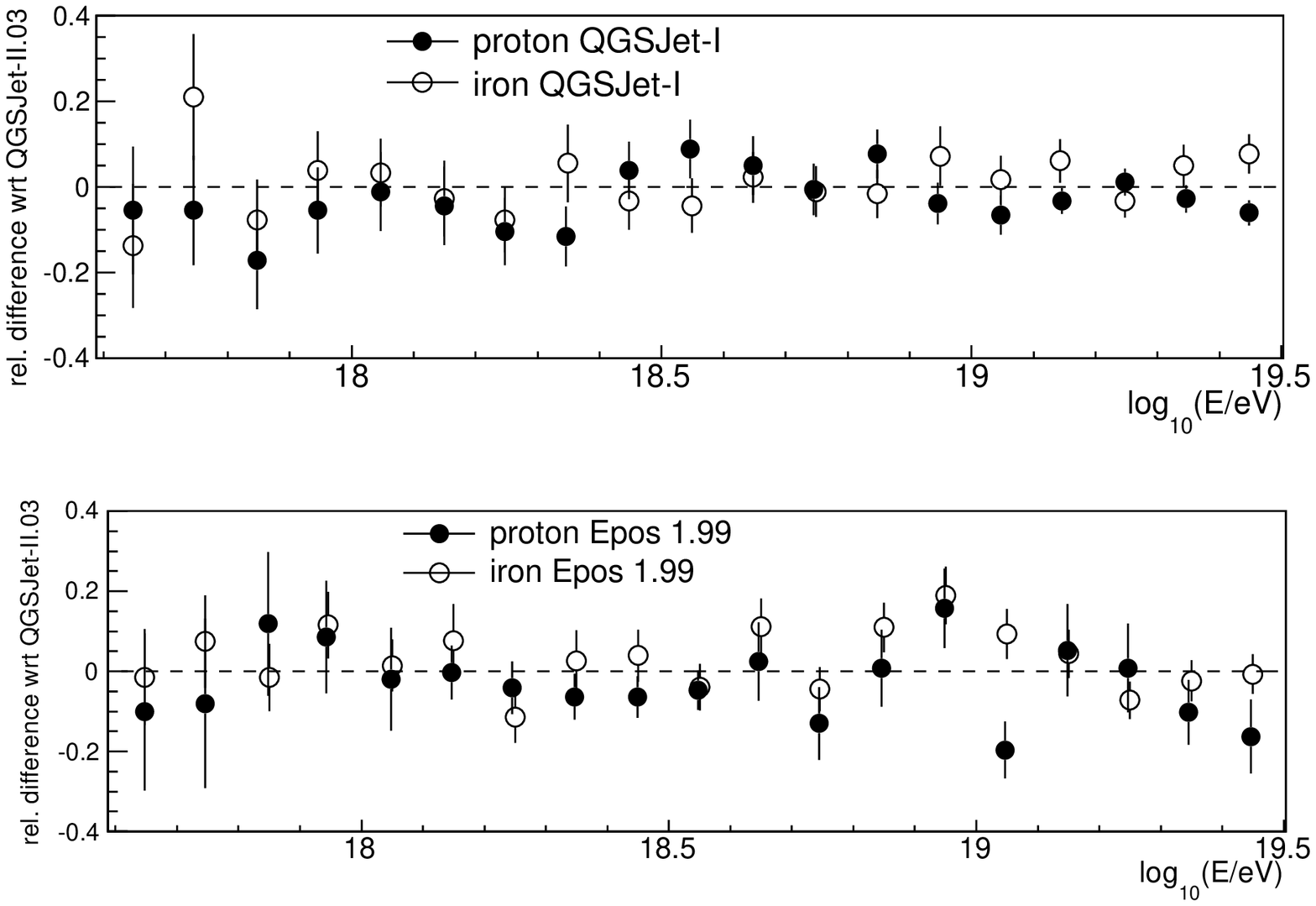}
\hspace{-0.4cm}
  \includegraphics[width=0.54\textwidth]{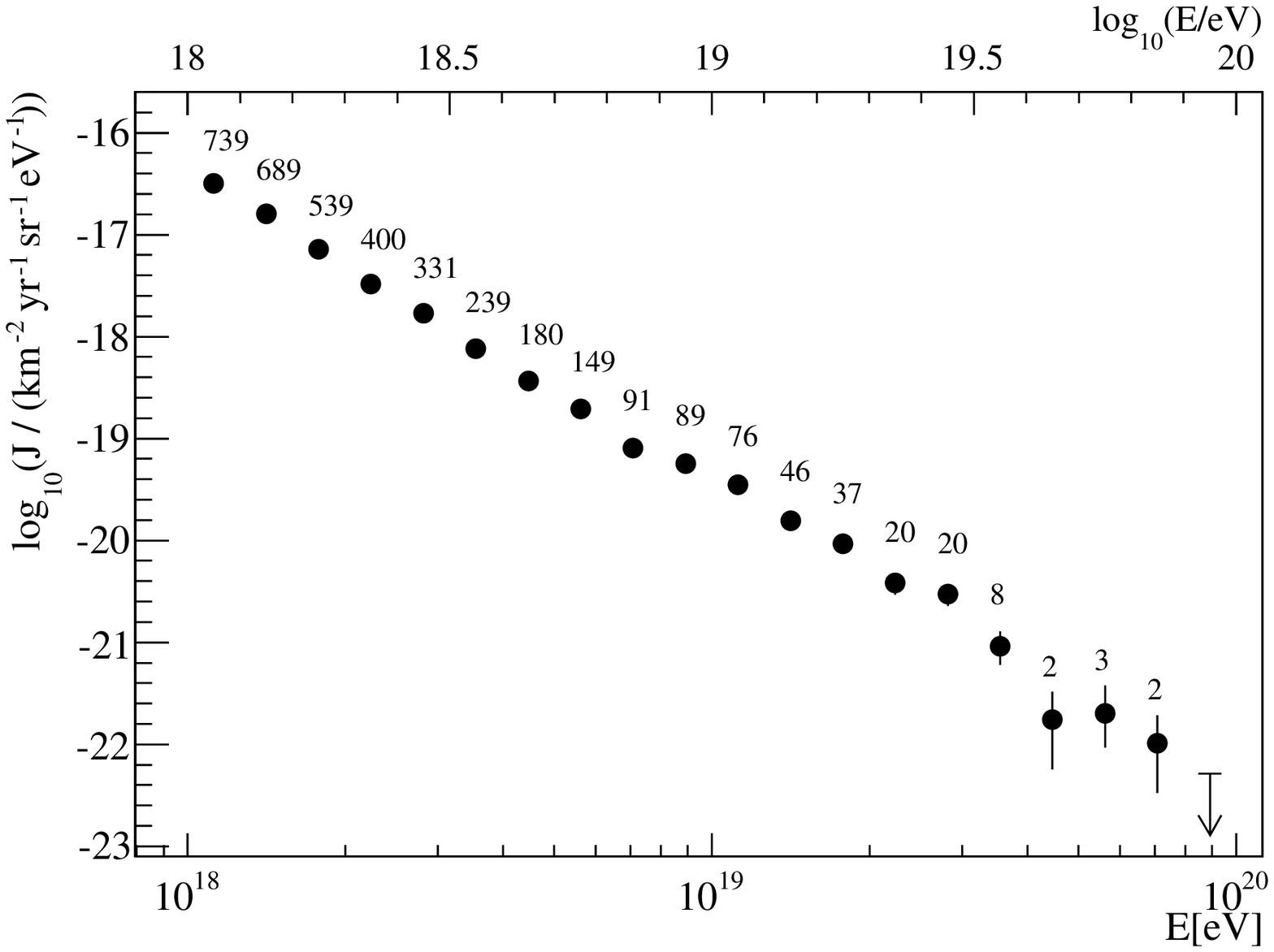}
  \caption{Left: relative differences, in exposure, between QGSJET-I (top) and Epos~1.99 (bottom) with respect to QGSJET-II.03, for proton (filled markers) and iron (empty markers). Right: energy spectrum derived from hybrid events using the standard method. Only statistical uncertainties are shown. Event numbers are indicated.}\label{fig:spectrum1}
\end{figure}

For the full Monte Carlo approach, the overall systematics are larger, dominated by the uncertainty on the mass composition. This is below than 10\% in the energy range above 10$^{18}$~eV. The systematic uncertainties on common tools (\emph{i.e.} on-time) and based on general cross-checks (laser shots and data/MC comparison) have been considered following the standard method. The dependence of the exposure on the hadronic interaction models has been checked using QGSJET-I and Epos-1.99 as additional models for the EAS generation. The relative difference of these models with respect to QGSJET-II, is shown up to 10$^{19.5}$~eV in Fig.~\ref{fig:spectrum1}, left. An impact of about 2\% has been found over the full energy range and assuming a mixed composition. Since the choice of different atmospheric profile may also influence the shower development, EAS simulations have been performed using realistic Malarg\"ue seasonal models and the US standard profile models, implemented in CORSIKA. The final impact on the exposure is smaller than 2\%.
Compared to the standard method, an additional contribution of less than 4\% (Fig.~\ref{fig:systematics}, left) has to be considered because of the looser FD fiducial distance cut used in the full Monte Carlo analysis (see section~\ref{sect:corsika}). This contribution includes the systematic uncertainty related to different choices of fluorescence yield that may change the maximum triggerable volumes at given energies. This check only refers to trigger and selection efficiency, since a consistent fluorescence yield is used in the simulation and the reconstruction phases. The impact of a different fluorescence yield on data reconstruction is included in the uncertainty on the energy scale and will be discussed in the next section. 
The overall systematic uncertainty in the exposure does not exceed $\sim$~13\% above 10$^{18}$~eV for the full Monte Carlo approach. A discussion of the energy region below 10$^{18}$~eV is given in the appendix. 

\section{Energy spectrum}\label{sect:spectrum}
\begin{figure}[!ht]
\begin{center}
\includegraphics[width=0.62\textwidth]{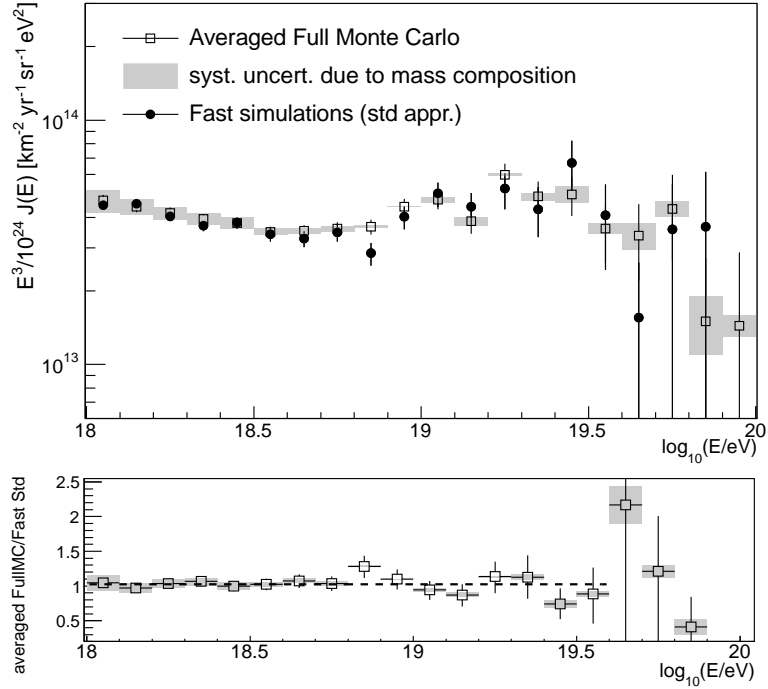}
\end{center}
 \caption{Comparison between the spectra measured using the hybrid exposure calculated with the fast simulation (dots) and the full Monte Carlo approach (empty squares). Because of the looser cuts used for the full Monte Carlo, an averaged spectrum is shown, with the systematic uncertainties due to mass composition (dark gray boxes). Bottom panel: ratio between the two spectra. }\label{fig:spectrumComp}
\end{figure}

The flux of cosmic rays $J$ as a function of energy is shown in Fig.~\ref{fig:spectrum1} (right) as derived using the standard method together with the number of selected events in each energy bin. In Fig.~\ref{fig:spectrumComp} this spectrum (dots) is compared to the one derived from the full Monte Carlo approach (empty squares with gray boxes). To emphasize their features, the two energy spectra are multiplied by an E$^3$ factor.
In the full Monte Carlo approach, because of the systematic uncertainties and the lack of knowledge of the mass composition of cosmic rays in this energy range, the spectrum has been derived using the two extreme assumptions of pure proton and pure iron composition. The missing energy~\cite{barbosa} assigned to data is chosen according to the primary mass. These two assumptions delimit a confidence region (gray boxes) in which the all-particle spectrum is expected to be found. 
The relative difference between the spectra derived with the two approaches is shown in the bottom panel. They differ by less than a few percent and they are compatible within the uncertainties. This good agreement between the two approaches, which are quite different concerning data sample, cuts and methods, is a nice confirmation for the resulting flux spectrum.

The main source of systematic uncertainty on the energy spectrum is the 22\%  on the energy scale. In particular, the largest contribution (∼14\%) is given by the absolute scale of the fluorescence yield~\cite{nagano}. The absolute calibration of the fluorescence telescope contributes about 9\%. An additional uncertainty of about 5\% is due to the measurement of atmospheric pressure, humidity and temperature and 4\%-8\% (depending on energy) is related to the attenuation of the light. Uncertainties of the lateral width of the shower image and other steps in the hybrid reconstruction method contribute about 9.5\% to the total uncertainty in the measured energy. 
The fraction of energy of the primary particle that is carried by muons and neutrinos has been calculated based on air shower simulations and goes from about 14\% at 10$^{18}$~eV to about 10\% at 10$^{19}$ eV. The systematic uncertainty depending on the choice of models and mass composition, is about 4\%. 
Indirect methods~\cite{castellina} of determining the energy scale, which do not involve the fluorescence detector calibration, seem to indicate an energy normalisation that is higher than the one used here by an amount comparable to the systematic uncertainty (22\%) given above.

\begin{figure}[!t]
\centering
 \resizebox{0.70\textwidth}{!}{\includegraphics{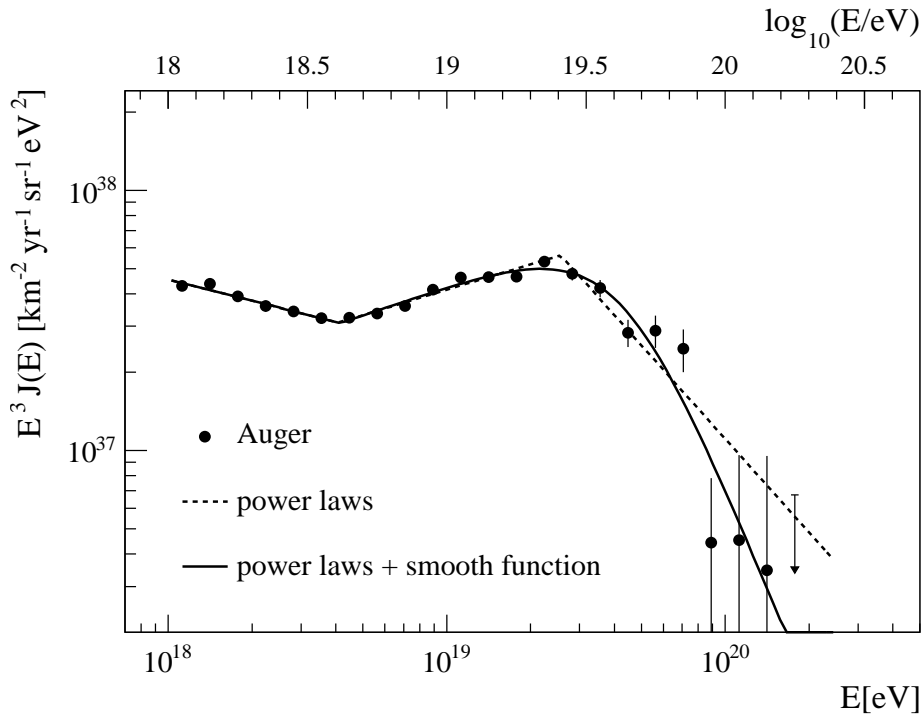}}
\caption{The energy spectrum obtained by combining the hybrid spectrum (standard approach) and the one measured with SD data. It is fitted with three power laws functions (dashed) and two power laws plus a smooth function (solid line). Only statistical uncertainties are shown. The systematic uncertainty on the energy scale is 22\%. }\label{fig:spectrum2}
\end{figure}

In Fig.~\ref{fig:spectrum2} the hybrid spectrum, derived with the standard approach, has been combined with the one measured from data collected by the surface detector above 10$^{18.5}$~eV.  Since the SD energy estimator is calibrated from a subset of high-quality hybrid events~\cite{SDspectrum,PesceICRC}, the two input spectra have the same systematic uncertainty of the energy scale while the flux normalisation uncertainties are independent. They are taken as 6\% for the SD and 10\%~(6\%) for the hybrid flux at 10$^{18}$~eV ($>$~10$^{19}$~eV). These normalisation uncertainties are used as additional constraints in the combination procedure which perform a maximum likelihood fit to derive the flux scaling factors {\it k}$_{\mathrm{SD}}$=1.01 and {\it k}$_{\mathrm{FD}}$=0.99 needed to match the two spectra.  

The characteristic features of the combined spectrum have been quantified with three power laws with free breaks between them (dashed line in Fig.~\ref{fig:spectrum2}) and with two power laws plus a smoothly changing function (solid line). The latter function is given by
\begin{eqnarray}
J(E;E>E_{\mathrm{ankle}})&\propto& E^{-\gamma_2} ~ \frac{1}{ 1 +
\exp\left(\frac{\lg{E}-\lg{E_{\frac{1}{2}}}}{\lg{W_c}}\right) },
\nonumber
\end{eqnarray}
where $E_{\frac{1}{2}}$ is the energy at which the flux has fallen to one half of the value of the power-law extrapolation and $W_c$ parametrizes the width of the transition region. 
The hypothesis that the power law above the ankle continues to highest energies with the spectral index $\gamma_2$ can be rejected with more than 20~$\sigma$.
 The derived parameters are given in Table~\ref{tab.1} quoting only the statistical uncertainties. The updated energy calibration curve~\cite{PesceICRC} has resulted in some changes of the parameters of the spectrum with respect to previous work, although only the values of $\gamma_2$ are different by more than the quoted statistical uncertainties~(values of $2.59\pm0.02$ and $2.55\pm0.04$ are reported in~\cite{spectrum2010} for $\gamma_2$ in the two cases of fit with three broken power laws and two power laws + smooth function respectively).

\begin{table}[!ht]
\begin{center}
\begin{tabular}{l|r|r}\hline
parameter & broken power laws & power laws \\
 & & + smooth function \\ \hline 
& & \\
$\gamma_1 (E < E_\mathrm{ankle})$ & $3.27 \pm 0.02$ & $3.27 \pm 0.01$
\\ 
$\lg(E_\mathrm{ankle}/\mathrm{eV})$ & $18.61 \pm 0.01$ & $18.62 \pm
0.01$ \\ 
$\gamma_2 (E > E_\mathrm{ankle})$ & $2.68 \pm 0.01$ &  $2.63 \pm 0.02$
\\ 
$\lg(E_\mathrm{break}/\mathrm{eV})$ & $19.41 \pm 0.02$ & \\ 
$\gamma_3 (E > E_\mathrm{break}) $ & $4.2 \pm 0.1$ &  \\ 
$\lg(E_\mathrm{\frac{1}{2}}/\mathrm{eV})$ &  & $19.63 \pm 0.02$\\ 
 $\lg(W_\mathrm{c}/\mathrm{eV})$ &  & $0.15 \pm 0.02$ \\ 
$\chi^2/\mathrm{ndof}$ & $37.8 / 16 = 2.4$ & $33.7 / 16 = 2.1$ \\ \hline
\end{tabular}
\end{center}
\caption{\label{tab.1}Fitted parameters and their statistical uncertainties characterizing the combined energy spectrum.}
\end{table}

\section{Summary}

The measurement of the cosmic ray flux above 10$^{18}$~eV has been updated to September 2010 using hybrid events of the Pierre Auger Observatory. The standard approach used here, and already adopted in a previous publication~\cite{spectrum2010}, is based on fast CONEX and detector simulations. In this paper the energy spectrum has additionally been derived using a full Monte Carlo method, based on CORSIKA air showers and detailed simulations of the hybrid detector. The full Monte Carlo approach provides a complete treatment of the shower-to-shower fluctuations, even in a region where the detector is not fully efficient and is an independent validation of the standard method. Producing a huge number of events is however computationally demanding. The lack of accurate knowledge of the mass composition propagates to the spectrum giving a confidence region for the expected flux. This is defined by the two extreme assumptions of pure proton and pure iron composition. 
Tighter cuts, designed to reduce this systematic uncertainty, are used in the standard method, profiting from the enormous statistics provided by the fast simulations. The average values of the spectra derived with the two approaches agree to within a few percent. 
In both cases, the dominant contribution to the systematic uncertainties in the flux measurement comes from the overall uncertainty on the energy scale, which is estimated to be 22\%.
The energy spectrum from the standard approach has been combined with the one derived above 10$^{18.5}$~eV by the surface array between January 2004 and December 2010. This updated combination provides an accurate determination of the spectral features in the energy range between 10$^{18}$~eV and 10$^{20}$~eV. The position of the ankle has been found to be at log$_{10}(\rm{E/eV})\,=\,18.61\,\pm\,0.01$ and a flux suppression has been observed at log$_{10}(\rm{E/eV})\,=\,19.63\,\pm\,0.02$, with a significance larger than 20~$\sigma$.

\section*{Acknowledgements}
The successful installation, commissioning, and operation of the Pierre Auger Observatory would not have been possible without the strong commitment and effort from the technical and administrative staff in Malarg\"ue.

We are very grateful to the following agencies and organizations for financial support: 
Comisi\'on Nacional de Energ\'ia At\'omica, 
Fundaci\'on Antorchas,
Gobierno De La Provincia de Mendoza, 
Municipalidad de Malarg\"ue,
NDM Holdings and Valle Las Le\~nas, in gratitude for their continuing
cooperation over land access, Argentina; 
the Australian Research Council;
Conselho Nacional de Desenvolvimento Cient\'ifico e Tecnol\'ogico (CNPq),
Financiadora de Estudos e Projetos (FINEP),
Funda\c{c}\~ao de Amparo \`a Pesquisa do Estado de Rio de Janeiro (FAPERJ),
Funda\c{c}\~ao de Amparo \`a Pesquisa do Estado de S\~ao Paulo (FAPESP),
Minist\'erio de Ci\^{e}ncia e Tecnologia (MCT), Brazil;
AVCR AV0Z10100502 and AV0Z10100522, GAAV KJB100100904, MSMT-CR LA08016,
LG11044, MEB111003, MSM0021620859, LA08015 and TACR TA01010517, Czech Republic;
Centre de Calcul IN2P3/CNRS, 
Centre National de la Recherche Scientifique (CNRS),
Conseil R\'egional Ile-de-France,
D\'epartement  Physique Nucl\'eaire et Corpusculaire (PNC-IN2P3/CNRS),
D\'epartement Sciences de l'Univers (SDU-INSU/CNRS), France;
Bundesministerium f\"ur Bildung und Forschung (BMBF),
Deutsche Forschungsgemeinschaft (DFG),
Finanzministerium Baden-W\"urttemberg,
Helmholtz-Gemeinschaft Deutscher Forschungszentren (HGF),
Ministerium f\"ur Wissenschaft und Forschung, Nordrhein-Westfalen,
Ministerium f\"ur Wissenschaft, Forschung und Kunst, Baden-W\"urttemberg, Germany; 
Istituto Nazionale di Fisica Nucleare (INFN),
Ministero dell'Istruzione, dell'Universit\`a e della Ricerca (MIUR), Italy;
Consejo Nacional de Ciencia y Tecnolog\'ia (CONACYT), Mexico;
Ministerie van Onderwijs, Cultuur en Wetenschap,
Nederlandse Organisatie voor Wetenschappelijk Onderzoek (NWO),
Stichting voor Fundamenteel Onderzoek der Materie (FOM), Netherlands;
Ministry of Science and Higher Education,
Grant Nos. N N202 200239 and N N202 207238, Poland;
Portuguese national funds and FEDER funds within COMPETE - Programa Operacional Factores de Competitividade through 
Funda\c{c}\~ao para a Ci\^{e}ncia e a Tecnologia, Portugal;
Ministry for Higher Education, Science, and Technology,
Slovenian Research Agency, Slovenia;
Comunidad de Madrid, 
Consejer\'ia de Educaci\'on de la Comunidad de Castilla La Mancha, 
FEDER funds, 
Ministerio de Ciencia e Innovaci\'on and Consolider-Ingenio 2010 (CPAN),
Xunta de Galicia, Spain;
Science and Technology Facilities Council, United Kingdom;
Department of Energy, Contract Nos. DE-AC02-07CH11359, DE-FR02-04ER41300,
National Science Foundation, Grant No. 0450696,
The Grainger Foundation USA; 
NAFOSTED, Vietnam;
Marie Curie-IRSES/EPLANET, European Particle Physics Latin American Network, 
European Union 7th Framework Program, Grant No. PIRSES-2009-GA-246806; 
and UNESCO.

\appendix
\section*{Appendix A: Systematic uncertainties below 10$^{18}$~eV}

\begin{figure}[hbt]
\centering
 \resizebox{0.6\textwidth}{!}{\includegraphics{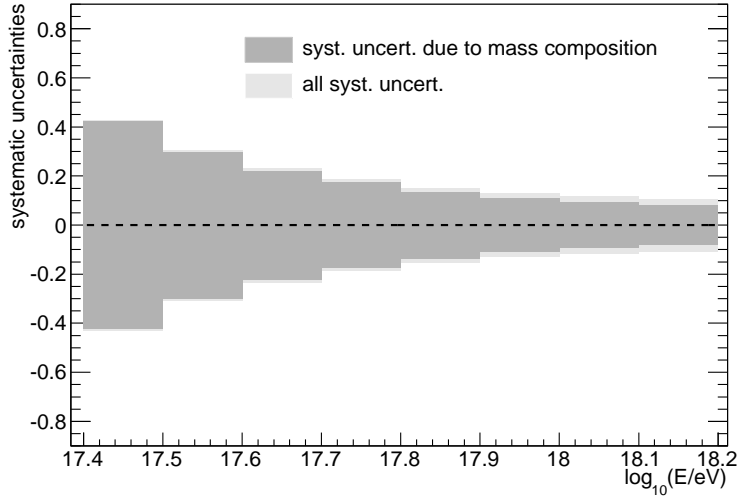}}
\caption{Systematic uncertainty expected on the energy spectrum below 10$^{18}$~eV due to mass composition (dark gray) and including all the other contributions (light gray) except for the one on the energy scale.}\label{fig:lowE}
\end{figure}

Given the increasing interest in the energy range below 10$^{18}$~eV, it is worthwhile to briefly discuss here the limitations of the hybrid detector in its ``regular'' design to assess this region. 
The energy spectrum presented in this paper has been derived above 10$^{18}$~eV, where the hybrid detector is fully efficient, independently of mass composition (Fig.~\ref{fig:Ontime}, right). This is an essential requirement for an accurate and unbiased measurement of the CR flux. 
However, below 10$^{18}$~eV, the efficiency of detecting an event as hybrid if the FD has been triggered, is still larger than 90\% while it rapidly goes down and depends significantly on the mass of the primary particle at energies smaller than 10$^{17.5}$~eV.
Unlike the fast simulations, that would have to rely on the LTP parameterizations to provide a mean detector response in a region where the efficiency is not unity, the full Monte Carlo method offers the possibility of also investigating the systematic uncertainty accurately below the EeV energy range since the SD response, including the shower-to-shower fluctuations, is entirely simulated. This enables us to safely extend the exposure calculation down to 10$^{17.4}$~eV. 
The expected systematics on the energy spectrum are shown in Fig.~\ref{fig:lowE} as a band delimited by the iron and proton assumptions. The light gray shaded area is obtained by summing in quadrature all the sources of systematic uncertainties (see section~\ref{sect:systematics}) except for the one on the energy scale. As a consequence of the large systematic uncertainty (about 30\% at 10$^{17.5}$~eV), a precise measurement of the all-particle CR flux with the regular hybrid detector is thus limited without a precise knowledge of the X$_{\rm{max}}$ distribution of CRs. It is worthwhile noting, that the systematic uncertainties shown in Fig.~\ref{fig:lowE} can be significantly reduced if fiducial field of view cuts as in~\cite{massAuger1} are applied.

An accurate measurement of the energy spectrum in this energy range will be possible with the detector enhancements (see section~\ref{sect:auger}) already installed at the Pierre Auger Observatory. In particular the dense array will allow us to extend the region of SD full detection efficiency down to about 10$^{17.5}$~eV~\cite{amiga2}. At the same time, the HEAT telescopes are enlarging the geometrical field of view thus reducing the current limitation on the measurement of the mass composition at lower energy.
\newpage
\section*{\small{The Pierre Auger Collaboration}}
{\scriptsize
P.~Abreu$^{63}$, 
M.~Aglietta$^{51}$, 
M.~Ahlers$^{94}$, 
E.J.~Ahn$^{81}$, 
I.F.M.~Albuquerque$^{15}$, 
D.~Allard$^{29}$, 
I.~Allekotte$^{1}$, 
J.~Allen$^{85}$, 
P.~Allison$^{87}$, 
A.~Almela$^{11,\: 7}$
% ...

\par\noindent
$^{1}$ Centro At\'{o}mico Bariloche and Instituto Balseiro (CNEA-UNCuyo-CONICET), San 
Carlos de Bariloche, 
Argentina. 
% ...
$^{7}$ Instituto de Tecnolog\'{\i}as en Detecci\'{o}n y Astropart\'{\i}culas (CNEA, CONICET, UNSAM), 
Buenos Aires, 
Argentina. 
% ...
$^{11}$ Universidad Tecnol\'{o}gica Nacional - Facultad Regional Buenos Aires, Buenos Aires,
Argentina. 
% ...
$^{15}$ Universidade de S\~{a}o Paulo, Instituto de F\'{\i}sica, S\~{a}o Paulo, SP, 
Brazil. 
% ...
$^{29}$ Laboratoire AstroParticule et Cosmologie (APC), Universit\'{e} Paris 7, CNRS-IN2P3, 
Paris, 
France. 
% ...
$^{51}$ Istituto di Fisica dello Spazio Interplanetario (INAF), Universit\`{a} di Torino and 
Sezione INFN, Torino, 
Italy. 
% ...
$^{63}$ LIP and Instituto Superior T\'{e}cnico, Technical University of Lisbon, 
Portugal. 
% ...
$^{81}$ Fermilab, Batavia, IL, 
USA. 
% ...
$^{85}$ New York University, New York, NY, 
USA. 
% ...
$^{87}$ Ohio State University, Columbus, OH, 
USA. 
% ...
$^{94}$ University of Wisconsin, Madison, WI, 
USA. 
% ...
Vietnam. 
\par\noindent
(\ddag) Deceased. 
(a) at Konan University, Kobe, Japan. 
(b) now at the Universidad Autonoma de Chiapas on leave of absence from Cinvestav. 
(f) now at University of Maryland. 
(h) now at NYU Abu Dhabi. 
(i) now at Universit\'{e} de Lausanne.
}
\end{document}